\documentstyle[11pt,fleqn]{article}

\begin{document}

\title{{\bf Factorisation of analytic representations in the unit disk
and number-phase statistics of a quantum harmonic oscillator}}
\author{ {\Large
A Vourdas,${}^{1}$\thanks{E-mail: ee21@liverpool.ac.uk}
\ C Brif$\,{}^{2}$\thanks{E-mail: costya@physics.technion.ac.il} 
\ and A Mann$\,{}^{2}$\thanks{E-mail: ady@physics.technion.ac.il} } \\
\\
{\normalsize ${}^{1}$Department of Electrical Engineering and 
Electronics, University of Liverpool, } \\
{\normalsize Brownlow Hill, Liverpool L69 3BX, United Kingdom } \\
{\normalsize ${}^{2}$Department of Physics, Technion -- 
Israel Institute of Technology, Haifa 32000, Israel } }
\date{}
\maketitle

\begin{abstract}
\noindent
The inner-outer part factorisation of analytic representations in the 
unit disk is used for an effective characterisation of the number-phase 
statistical properties of a quantum harmonic oscillator. 
It is shown that the factorisation is intimately connected to the 
number-phase Weyl semigroup and its properties.
In the Barut-Girardello analytic representation the factorisation is
implemented as a convolution.
Several examples are given which demonstrate the physical significance 
of the factorisation and its role for quantum statistics.
In particular, we study the effect of phase-space interference 
on the factorisation properties of a superposition state. 
\end{abstract} 

\vspace*{0.5cm}

\section{Introduction}

The powerful theory of analytic functions has proved to be very useful
in many areas of theoretical physics. Here we use this theory to study the  
number-phase statistics of a quantum harmonic oscillator.
This approach employs the inner-outer part factorisation of analytic 
representations in the unit disk \cite{Duren,Hoff}. This representation is 
based on the Perelomov SU(1,1) coherent states \cite{Per}. 
In the present work we consider the Holstein-Primakoff   
realization of the SU(1,1) Lie algebra \cite{HoPr,Ger83}, where 
the SU(1,1) generators act on the Hilbert space of a quantum 
harmonic oscillator (the Fock space) spanned by the number states 
$|n\rangle$ $(n=0,1,2,\ldots)$. 
The SU(1,1) coherent states in this case are the eigenstates of the 
exponential phase operator \cite{SG,CN} and are closely related to the 
problem of the quantum description of oscillator phase
\cite{Lern,Ifant,Ahar,LL,Vou90,Vou92,Vou93,Brif95}.
(For the present status of the quantum phase problem see e.g. 
\cite{Scripta,Ly,Tanas}.)
Some other aspects of the SU(1,1) coherent states were studied in
\cite{Kur68,Sud93}.

In section \ref{sec2} we introduce the analytic representations in the 
unit disk and their `boundaries' which are the phase-state representations 
on the unit circle \cite{GW,BBA94b}. In section \ref{sec3} we discuss the 
factorisation of these analytic representations into a product of an 
inner and an outer function. Important information about a quantum
state can be inferred from the factorisation of the corresponding 
analytic function. For example, the phase distribution uniquely 
determines the outer part, and vice versa. The SU(1,1) coherent states
and the Barut-Girardello states \cite{BG} are outer states, i.e. their 
analytic representations are outer functions. All physical properties of 
such states are uniquely determined by their phase distribution. 
In section \ref{sec4} we demonstrate that although the factorisation 
is intimately connected to the analytic representations in the unit disk,
it can also be implemented in other representations. We consider the 
Barut-Girardello analytic representation \cite{BG} and show that there 
the factorisation becomes convolution.
In section \ref{sec5} we give several examples. It is shown that 
interference in phase space between components of a quantum superposition  
state has important effects on the factorisation of the corresponding 
analytic function. We construct explicitly a ``Blaschke state'' whose 
analytic representation in the unit disc is a Blaschke factor. It becomes 
clear that many aspects of the mathematical theory of Hardy spaces have a 
physical meaning.
In section \ref{sec6} we demonstrate the connection between the 
factorisation and the phase-space formalism based on the number and phase 
variables. We consider ``shifted states''  and the number-phase 
Wigner functions.
We conclude in section \ref{sec:concl} with discussion of our results.

\section{Analytic representations in the unit disk}
\label{sec2}

It is known that the SU(1,1) generators $\hat{K}_{\pm}$ and $\hat{K}_{0}$ 
can be realized in the Fock space of a quantum harmonic oscillator.
For Bargmann index equal to one half, this realization reads 
\cite{HoPr,Ger83}
 	\begin{equation}
\hat{K}_{+} = \hat{N}^{1/2} \hat{a}^{\dagger}  \;\;\;\;\;\;\;\;\;\;
\hat{K}_{-} = \hat{a} \hat{N}^{1/2} \;\;\;\;\;\;\;\;\;\;
\hat{K}_{0} = \hat{N} + \mbox{$\frac{1}{2}$} 
\label{2.1}
	\end{equation}
where $\hat{a}$ and $\hat{a}^{\dagger}$ are the usual annihilation and
creation operators and  $\hat{N} = \hat{a}^{\dagger}\hat{a}$ is the number 
operator. Then the Perelomov SU(1,1) coherent states are given by
 	\begin{eqnarray}
& & |z\rangle = \exp( \xi\hat{K}_{+} - \xi^{\ast}\hat{K}_{-} ) |0\rangle
= (1-|z|^{2})^{1/2} \sum_{n=0}^{\infty} z^n |n\rangle  \label{2.2} \\
& & z = (\xi/|\xi|)\tanh |\xi| \;\;\;\;\;\;\;\;\;\; |z|<1 . \label{2.3}
	\end{eqnarray}
These states form an overcomplete basis in the Hilbert space \cite{Per}.
They are the eigenstates of the exponential phase operator,
	\begin{equation}
\hat{E}_{-} |z\rangle = z |z\rangle    \label{2.4}
	\end{equation}
and in this sense they have also been called phase states 
\cite{Vou90,Vou92,Vou93}. Some interesting properties of the states 
$|z\rangle$ were discussed in \cite{Kur68,Sud93}.
The exponential phase operators \cite{SG,CN}
 	\begin{eqnarray}
& & \hat{E}_{-} = \sum_{n=0}^{\infty} |n\rangle \langle n+1|
\;\;\;\;\;\;\;\;\;\; \hat{E}_{+} = \hat{E}_{-}^{\dagger} = 
\sum_{n=0}^{\infty} |n+1\rangle \langle n|
\label{2.5} \\
& & \hat{E}_{-} |0\rangle = 0 \;\;\;\;\;\;\;\;\;\;
\hat{E}_{-} \hat{E}_{+} = \hat{1} \;\;\;\;\;\;\;\;\;\;
\hat{E}_{+} \hat{E}_{-} = \hat{1} - |0\rangle \langle 0|  \label{2.6}
	\end{eqnarray}
are related to $\hat{a}, \hat{a}^{\dagger}$ through the polar 
decomposition:
	\begin{equation}
\hat{a} = \hat{E}_{-} { \hat{N} }^{1/2} \;\;\;\;\;\;\;\;\;\;
\hat{a}^{\dagger} = { \hat{N} }^{1/2} \hat{E}_{+} .  \label{2.7}
	\end{equation}
With these definitions, the Holstein-Primakoff realization (\ref{2.1}) 
takes the form
	\begin{equation}
\hat{K}_{+} = \hat{N} \hat{E}_{+}  \;\;\;\;\;\;\;\;\;\;
\hat{K}_{-} = \hat{E}_{-} \hat{N}   \;\;\;\;\;\;\;\;\;\;
\hat{K}_{0} = \hat{N} + \mbox{$\frac{1}{2}$} .
\label{2.8}
	\end{equation}

Let $|f\rangle$ be an arbitrary (normalized) state
	\begin{equation}
|f\rangle = \sum_{n=0}^{\infty} f_{n} |n\rangle \;\;\;\;\;\;\;\;\;\;
\sum_{n=0}^{\infty} |f_{n}|^{2} = 1 .   \label{2.9}
	\end{equation}
Its analytic representation in the unit disk $D(|z|<1)$ is the function
	\begin{equation}
Z(f;z) = (1-|z|^{2})^{-1/2} \langle f|z \rangle = \sum_{n=0}^{\infty} 
f_{n}^{\ast} z^n   \label{2.10}
	\end{equation}
which belongs to the Hardy space $H_{2}(D)$ \cite{Duren,Hoff}. Various 
properties and applications of this representation were considered in a 
number of works 
\cite{Vou92,Vou93,Kur68,Sud93,Vou_JMP,BBA94d,BBA96,last}.
For later use we point out that the number states $|n\rangle$ and the 
SU(1,1) coherent states $|z_0 \rangle$ are represented respectively
by the functions
	\begin{eqnarray}
& & Z(n;z) = z^{n} \label{2.11} \\
& & Z(z_{0};z) = \frac{ (1-|z_0|^2)^{1/2} }{ 1 - z_0^{\ast} z } .
\label{2.12}
	\end{eqnarray}

It is often convenient to use the number and phase representations.
One can introduce a `phase space' based on the number and phase variables 
and the number-phase Wigner function \cite{VP90,LP,Vacc}.
The number representation of the state $|f\rangle$ given by (\ref{2.9}) is
the sequence $\{f_{n} = \langle n|f \rangle \}$. The action of the number 
operator $\hat{N}$ is represented by multiplication by $n$, while the 
ladder operator $\hat{E}_{+}$ shifts the sequence by one place:
	\begin{equation}
\hat{N} f_{n} = n f_{n}    \;\;\;\;\;\;\;\;\;\;
\hat{E}_+ f_{n}= f_{n - 1} .      \label{2.13}
	\end{equation}

The phase representation is based on the phase states 
	\begin{equation}
|\theta\rangle = \lim_{|z| \rightarrow 1} (1-|z|^2)^{-1/2} 
\left| z=|z|{\rm e}^{ {\rm i} \theta} \right\rangle = \sum_{n=0}^{\infty}
{\rm e}^{ {\rm i} n\theta} |n\rangle .   \label{2.14}
	\end{equation}
According to (\ref{2.4}), one has 
$\hat{E}_{-} |\theta\rangle = {\rm e}^{ {\rm i} \theta} |\theta\rangle$.
The $|\theta\rangle$ states are unnormalizable, nonorthogonal and 
formally they do not belong to the Hilbert space. However, they do 
resolve the identity,
	\begin{equation}
\frac{1}{2\pi} \int_{-\pi}^{\pi} {\rm d} \theta \, 
|\theta\rangle\langle\theta| = \hat{1}   \label{2.15}
	\end{equation}
and the phase-state representation is a useful calculational tool
\cite{GW,BBA94b}. The normalized state $|f\rangle$ of the form (\ref{2.9})
is represented by the periodic function
	\begin{equation}
\Theta(f;\theta) = \langle f|\theta \rangle = \lim_{|z| \rightarrow 1} 
Z(f; z=|z|{\rm e}^{ {\rm i} \theta}) = \sum_{n=0}^{\infty} f_{n}^{\ast}
{\rm e}^{ {\rm i} n\theta}                     \label{2.16}
	\end{equation}
which is the ``boundary function'' of the function $Z(f;z)$.
For the sake of simplicity we denote this function as $\Theta(f;\theta)$
instead of $\Theta(f; {\rm e}^{ {\rm i} \theta})$ which would be a more 
appropriate notation. The limit $|z| \rightarrow 1$ in (\ref{2.16}) exists
for any normalizable state $|f\rangle$. The definition of $\Theta(f;\theta)$ 
as the boundary function of $Z(f;z)$ is mathematically more rigorous, 
while the definition as $\langle f|\theta \rangle$ is physically more 
appealing. By using the identity resolution (\ref{2.15}), one finds
	\begin{eqnarray}
& & |f\rangle = \frac{1}{2\pi} \int_{-\pi}^{\pi} {\rm d} \theta \, 
\Theta^{\ast}(f;\theta) |\theta\rangle \label{2.17} \\
& & \langle g|f \rangle = \frac{1}{2\pi} \int_{-\pi}^{\pi} {\rm d} \theta \,
\Theta(g;\theta) \Theta^{\ast}(f;\theta) .  \label{2.18}
	\end{eqnarray}
It is easily seen that the boundary function $\Theta(f;\theta)$ 
determines uniquely the analytic function $Z(f;z)$:
	\begin{eqnarray}
& & Z(f; z=r {\rm e}^{ {\rm i} \theta}) = \frac{1}{2\pi} 
\int_{-\pi}^{\pi} {\rm d} \theta' \, C(r,\theta-\theta') 
\Theta(f;\theta')   \label{2.19} \\
& & C(r,\theta) \equiv ( 1-r {\rm e}^{ {\rm i} \theta} )^{-1} . 
\label{2.20}
	\end{eqnarray}
The number operator $\hat{N}$ and the ladder operator $\hat{E}_{+}$ act 
in the phase representation as
	\begin{equation}
\hat{N} \Theta(f;\theta) = - {\rm i} \frac{ \partial }{ \partial \theta }
\Theta(f;\theta) \;\;\;\;\;\;\;\;\;\;
\hat{E}_+ \Theta(f;\theta) = \exp( {\rm i} \theta) \Theta(f;\theta) .
\label{2.21}
	\end{equation}

The phase properties of a state are determined by its phase distribution
\cite{SS,VaPe,BBA94a,Leo}.
	\begin{equation}
P(f;\theta) = \frac{1}{2\pi} |\Theta(f;\theta)|^{2} = \frac{1}{2\pi}
\sum_{n,m=0}^{\infty} f_{n}^{\ast} f_{m} {\rm e}^{ {\rm i} (n-m) \theta} .
\label{2.22}
	\end{equation}
This is a positive periodic function of $\theta$ normalized by
$\int_{2\pi} {\rm d} \theta \, P(f;\theta) = 1$.
The number states $|n\rangle$ have the uniform phase distribution 
$P(n;\theta) = (2\pi)^{-1}$. For the SU(1,1) coherent states 
 $|z\rangle$ one obtains
$P(z;\theta) = (2\pi)^{-1} {\cal P}(r,\theta-\phi)$, where 
$z=r {\rm e}^{ {\rm i} \phi}$ and
	\begin{equation}
{\cal P}(r,\theta) = {\rm Re}\, [ 2C(r,\theta) -1] = 
\frac{ 1-r^2 }{ 1+r^2-2r \cos\theta }   \label{2.23}
	\end{equation}
is the Poisson kernel. For later use we also define the harmonic 
conjugate of ${\cal P}(r,\theta)$:
	\begin{equation}
{\cal Q}(r,\theta) = {\rm Im}\, [ 2C(r,\theta) -1] = 
\frac{ 2r \sin\theta }{1+r^2-2r \cos\theta } .  \label{2.24}
	\end{equation}
Note the following limits:
	\begin{equation}
\lim_{r \rightarrow 1} {\cal P}(r,\theta) = 2\pi \delta(\theta) 
\;\;\;\;\;\;\;\;\;\;
\lim_{r \rightarrow 1} {\cal Q}(r,\theta) = \cot(\theta/2) .   
\label{2.25}
	\end{equation}

The Weyl semigroup for the number and phase operators is introduced as
\cite{Vou92}
	\begin{eqnarray}
& & \hat{W}(m,\beta,\gamma) = \hat{E}_{+}^{m} 
\exp( {\rm i} \beta \hat{N} )
\exp( {\rm i} \gamma )   \label{2.26} \\
& & \hat{W}(m_1 , \beta_1 , \gamma_1 ) 
\hat{W}(m_2 , \beta_2 , \gamma_2 )
= \hat{W}(m_1 + m_2 , \beta_1 + \beta_2 , \gamma_1 + 
\gamma_2 + m_2 \beta_1 )       \label{2.27} \\
& & \hat{W}(0,0,0) = \hat{1}    \;\;\;\;\;\;\;\;\;\;
\hat{W}^{\dagger} \hat{W} = \hat{1} \;\;\;\;\;\;\;\;\;\;
\hat{W} \hat{W}^{\dagger} = \hat{1} - \sum_{n=0}^{m-1} 
| n\rangle\langle n|   \label{2.28}
	\end{eqnarray}
where $\beta$ and $\gamma$ are real parameters and $m$ is a non-negative 
integer. The operators $\hat{W}$ are isometric but not unitary. 
They do not have an inverse and therefore they form a semigroup but
not a group. (Note that this semigroup becomes effectively a group
when the antinormal ordering of the phase operators is applied 
\cite{LP,BBA94a,VaBA}.) The operators $\hat{W}(m,\beta,\gamma)$
play a role similar to that of the Weyl displacement operators 
$\exp[ i (\mu \hat{x} + \nu \hat{p})]$ in the position-momentum 
phase space. For example, we mention the following basic properties:
	\begin{eqnarray}
& & \hat{E}_+ ^{m} |n\rangle = |n + m\rangle \;\;\;\;\;\;\;\;\;\; 
\exp( {\rm i} \beta \hat{N} ) |n\rangle = \exp( {\rm i} \beta n ) 
|n\rangle   \label{2.29} \\
& &  \hat{E}_{-}^{m} |z\rangle = z^m |z\rangle \;\;\;\;\;\;\;\;\;\; 
\exp( {\rm i} \beta \hat{N} ) |z\rangle = 
| z {\rm e}^{ {\rm i} \beta} \rangle .  \label{2.30}
	\end{eqnarray}

\section{Factorisation of analytic representations  
in the unit disk into their inner and outer parts}
\label{sec3}

We have explained in the previous section that an arbitrary state can be
represented in the Hardy space $H_{2}(D)$ of analytic functions in the
unit disk. An important tool in the theory of these functions is their
factorisation into the product of an inner and an outer function.
An analytic function $Z(f;z)$ can be expressed as \cite{Duren,Hoff}
	\begin{eqnarray}
& & Z(f;z) = Z_{{\rm in}}(f;z)  Z_{{\rm out}}(f;z) \label{3.1} \\
& & Z_{{\rm out}}(f;z) = \exp[ \Phi(f;z) ]  \label{3.2} \\
& & \Phi(f; z= r {\rm e}^{ {\rm i} \theta}) =\Phi_{R}(f;z) + {\rm i} 
\Phi_{I}(f;z) \nonumber \\ & & \mbox{\hspace{2.35cm}}
= \frac{1}{2\pi} \int_{-\pi}^{\pi} {\rm d} \theta' \, 
[2 C(r,\theta-\theta') - 1 ] \ln |\Theta(f;\theta')|    \label{3.3} \\
& & \Phi_{R}(f;z=r {\rm e}^{ {\rm i} \theta}) = \frac{1}{2\pi} 
\int_{-\pi}^{\pi} {\rm d} \theta' \, {\cal P}(r,\theta-\theta') 
\ln|\Theta(f;\theta')| \label{3.4} \\
& & \Phi_{I}(f;z=r {\rm e}^{ {\rm i} \theta}) = \frac{1}{2\pi} 
\int_{-\pi}^{\pi} {\rm d} \theta' \, {\cal Q}(r,\theta-\theta') 
\ln|\Theta(f;\theta')| \label{3.5} \\
& & Z_{{\rm in}}(f;z) = Z(f;z)/Z_{{\rm out}}(f;z) = 
Z(f;z) \exp[ -\Phi(f;z) ] .  \label{3.6}
	\end{eqnarray}
The functions $ Z_{{\rm in}}(f;z)$ and $Z_{{\rm out}}(f;z)$ are called
the inner and outer parts of $Z(f;z)$. It can be proved that in the
interior of the unit disk ($|z|<1$) the absolute value of the inner part
is bounded by 1:
	\begin{equation}
|Z_{{\rm in}}(f;z)| = |Z(f;z) \exp[ -\Phi(f;z) ]| \leq 1   \label{3.7}
	\end{equation}
and that on the unit circle ($|z|=1$) this absolute value is equal 1:
	\begin{equation}
|\Theta_{{\rm in}}(f;\theta)| = |\Theta(f;\theta) \exp[ -\Phi(f;\theta) ]| 
= |\Theta(f;\theta)| \exp[ -\Phi_{R}(f;\theta) ] = 1 .  \label{3.8}
	\end{equation}
We use the notation $\Theta(f;\theta)$, $\Theta_{{\rm in}}(f;\theta)$,
$\Phi(f;\theta)$, etc. for the boundary functions of $Z(f;z)$, 
$Z_{{\rm in}}(f;z)$, $\Phi(f;z)$, etc. obtained in the limit 
$|z| \rightarrow 1$ (with $z=|z| {\rm e}^{ {\rm i} \theta}$). 
Relations (\ref{3.7}) 
and (\ref{3.8}) form a necessary and sufficient condition for a function
$Z_{{\rm in}}(f;z)$ to be an inner function. 
It follows from equation (\ref{3.8}) that the phase distribution
	\begin{equation}
P(f;\theta) = \frac{1}{2\pi} |\Theta(f;\theta)|^2 =
\frac{1}{2\pi} |\Theta_{{\rm out}}(f;\theta)|^2 = \frac{1}{2\pi} 
\exp[ 2 \Phi_{R}(f;\theta) ]    \label{3.9}
	\end{equation}
depends only on the outer part of the function and more specifically
on $\Phi_{R}(f;\theta)$. Conversely, equation (\ref{3.3}) shows that
the $P(f;\theta)$ defines uniquely the outer part $Z_{{\rm out}}(f;z)$.  

The functions $\Phi_{R}(f;z)$, $\Phi_{I}(f;z)$ are real harmonic
functions and they are harmonic conjugate of each other. The function
$\ln |Z(f;z)|$ is a subharmonic function. Indeed, its boundary function
is
	\begin{equation}
\ln |\Theta(f;\theta)| = \ln |\exp[ \Phi(f;\theta) ]| = \Phi_{R}(f;\theta)
= \mbox{$\frac{1}{2}$} \ln [ 2\pi P(f;\theta) ]    \label{3.10}
	\end{equation}
while in the interior of the unit disk
	\begin{equation}
\ln |Z(f;z)| = \ln |Z_{{\rm in}}(f;z)| + \ln |Z_{{\rm out}}(f;z)|
\leq \ln |Z_{{\rm out}}(f;z)| = \Phi_{R}(f;z) .  \label{3.11}
	\end{equation}
A function which is identical with its inner (outer) part is called an 
inner (outer) function and the corresponding quantum state is called an 
inner (outer) state. It is interesting that all the properties of an 
outer state are determined by its phase distribution. Obviously, 
any two states with the same outer part have identical phase properties. 
For example, any inner state has the uniform phase distribution 
$(2\pi)^{-1}$. The number states $|n\rangle$ represented by the 
functions $z^{n}$ are inner states. An easy way to check whether a state 
is outer is to calculate the outer part $Z_{{\rm out}}(f;z)$ from 
equations (\ref{3.2}), (\ref{3.3}) and to compare it with the whole 
function $Z(f;z)$. However, there is a simpler criterion for outer 
functions. It can be proved \cite{Duren,Hoff} that a function 
$\exp[ \Phi(z) ]$ is an outer function if and only if
	 \begin{eqnarray}
\Phi_{R}(z=0) = \ln |\exp[ \Phi(z=0) ]| & = & 
\frac{1}{2\pi} \int_{-\pi}^{\pi} {\rm d} \theta \, 
\ln|\exp[ \Phi(\theta) ]| 
\nonumber \\ & = & \frac{1}{2\pi} \int_{-\pi}^{\pi}
{\rm d} \theta \, \Phi_{R}(\theta) = 
\frac{1}{4\pi}\int_{-\pi}^{\pi} {\rm d} 
\theta \, \ln [ 2\pi P(\theta) ] .    \label{3.12}
	 \end{eqnarray}
By using this theorem, equation (\ref{2.12}) and the integral
	 \begin{equation}
\frac{1}{2\pi} \int_{-\pi}^{\pi} {\rm d} \theta \, 
\ln \left| {\rm e}^{ {\rm i} \theta } - \frac{1}{z_0} \right| 
= - \ln | z_0 |  \;\;\;\;\;\;\;\;\;\;  | z_0 | < 1     \label{3.13}
	 \end{equation}
we can prove that the SU(1,1) coherent states $| z_0 \rangle$ are outer 
states. One can also check, by using the generating function for the 
Gegenbauer polynomials \cite{AS}, that the straightforward evaluation of 
the integral in (\ref{3.3}) yields 
$Z( z_0 ; z) = Z_{{\rm out}}( z_0 ;z)$.

\section{Factorisation in the Barut-Girardello representation}
\label{sec4}

The factorisation into inner and outer parts has been introduced in the 
context of analytic representations in the unit disk. In this section we 
discuss how this factorisation can be generalized for the Barut-Girardello 
representation \cite{BG}. Although the factorisation is intimately 
connected with the analytic representations in the unit disk, it is nice 
to demonstrate that it can also be implemented  
in other representations. We choose the Barut-Girardello representation 
because there is a simple transformation that connects it with the 
representation in the unit disk. The Barut-Girardello states are defined 
as the eigenstates of the SU(1,1) lowering generator $\hat{K}_{-}$ 
\cite{BG}:
	 \begin{equation}
\hat{K}_{-} |u\rangle = u |u\rangle .   \label{3.14}
	 \end{equation}   
Here $u$ is an arbitrary complex number. 
Various properties of these states were studied in different physical
contexts \cite{Brif95,Dod74,Agar88,Buz90,BBA94c}.
In the Holstein-Primakoff
realization (\ref{2.1}) the Barut-Girardello  states are given by 
	 \begin{equation}
|u\rangle = [I_{0}(2|u|)]^{-1/2} \sum_{n=0}^{\infty} \frac{ u^n }{n!}
|n\rangle    \label{3.15}
	 \end{equation}
where $I_{0}(x)$ is the zero-order Bessel function of the first kind.
The analytic representation of these states in the unit disk is
	 \begin{equation}
Z(u;z) = ( 1-|z|^2 )^{-1/2} \langle u|z \rangle = [I_{0}(2|u|)]^{-1/2}
\exp( u^{\ast} z ) .       \label{3.16} 
	 \end{equation}
By using the relation (\ref{3.12}) or by the straightforward evaluation 
of the integral in (\ref{3.3}), we can prove that the Barut-Girardello 
states $|u\rangle$ are outer states. The states $|u\rangle$ form an 
overcomplete set in the Fock space with the identity resolution
	 \begin{equation}
\int {\rm d} \mu(u) |u \rangle\langle u| = \hat{1} \;\;\;\;\;\;\;\;\;\;
{\rm d} \mu(u) = \frac{2}{\pi} K_{0}(2|u|) I_{0}(2|u|) {\rm d}^{2} u . 
\label{3.17} 
	 \end{equation}
The integration in (\ref{3.17}) is over the whole $u$ plane, and 
$K_{0}(x)$ is the zero-order Bessel function of the second kind.
One can define the Barut-Girardello analytic representation in 
the complex plane \cite{BG} (which is different from the familiar 
Bargmann representation \cite{Barg61}). 
A normalized state $|f\rangle$ is represented by the function 
	\begin{eqnarray}
& & U(f;u) = [I_{0}(2|u|)]^{1/2} \langle f|u \rangle = \sum_{n=0}^{\infty}
f_{n}^{\ast} \frac{ u^n }{n!} \label{3.18} \\
& & |f\rangle = \int {\rm d} \mu(u) [I_{0}(2|u|)]^{-1/2} U^{\ast}(f;u) 
|u\rangle .    \label{3.19}
	\end{eqnarray}
It can be shown \cite{last} that the Barut-Girardello representation 
and the representation in the unit disk are related through the 
``${\cal L}$ transformation'' (which is effectively the Laplace transform):
	\begin{equation}
Z(f;z) =  \frac{1}{z}\, {\cal L} [U(f;u)] = \frac{1}{z}
\int\limits _{0}^{\infty} {\rm d} u \,  U(f;u) \exp(- u/z)  \label{3.20} 
\end{equation}
where the integration is along the positive real semiaxis and $z$ 
belongs to the right half of the unit disk $({\rm Re}\, z > 0)$. 
This equation defines $Z(f;z)$ in the right half of the unit disk and 
through analytic continuation in the whole unit disk. The inverse 
transformation is
\begin{equation}
U(f;u) = {\cal L}^{-1} [z Z(f;z)] =
\frac{1}{ 2\pi {\rm i} } \int_{1 - {\rm i} \infty}^{1 + {\rm i} \infty}  
{\rm d} w\, \frac{ Z(f;1/w) }{w} \exp(wu)  .  \label{3.21}
	\end{equation}
The integration is along the line $1+ {\rm i} t$ where $t$ is a 
real number. If we define the functions
	\begin{eqnarray}
& & U_{{\rm in}}(f;u) = {\cal L}^{-1} 
[z Z_{{\rm in}}(f;z)] \\
& & U_{{\rm out}}(f;u) = {\cal L}^{-1} 
[Z_{{\rm out}}(f;z)]  \label{3.22}
	\end{eqnarray}
the substitution of equation (\ref{3.1}) into (\ref{3.21}) defines the 
factorisation in the context of the Barut-Girardello representation.
Note that the factorisation is here a convolution:
	\begin{equation}
U(f;u) = \int_{0}^{u} {\rm d} x \, U_{{\rm in}}(f;x) 
U_{{\rm out}}(f;u-x) .    \label{3.23}
	\end{equation}
The integration is along the line from 0 to the complex number $u$.
As an example we consider the number states $|n\rangle$ which are
inner states with $Z_{{\rm out}}(n;z) = 1$, $Z_{{\rm in}}(n;z) = z^n$. 
Then equations (\ref{3.21})--(\ref{3.22}) give
	\begin{equation}
U_{{\rm out}}(n;u) = 2 \delta(u) \;\;\;\;\;\;\;\;\;\;\;\;
U(n;u) = U_{{\rm in}}(n;u) = \frac{u^{n}}{n!} . \label{3.24}
	\end{equation}
(We use the convention in which the delta function $\delta(u)$ is
symmetrical about $u = 0$.)
For an outer state $|f\rangle$ with $Z(f;z) = Z_{{\rm out}}(f;z)$, 
we obtain
	\begin{equation}
U_{{\rm in}}(f;u) = 1 \;\;\;\;\;\;\;\;\;\;\;\;
U(f;u) = \int_{0}^{u} {\rm d} x \, U_{{\rm out}}(f;x) .  \label{3.25}
	\end{equation}
Then the outer part of the Barut-Girardello representation can be 
written as
	\begin{equation}
U_{{\rm out}}(f;u) = 2 f_{0}^{\ast} \delta(u) + 
\partial U(f;u)/\partial u .   \label{3.26}
	\end{equation}

\section{Examples: Quantum superposition states}
\label{sec5}

Quantum superposition states have many interesting properties such as 
squeezing and sub-Poissonian photon statistics \cite{BuKn}. 
These features arise because components of a 
superposition state interfere with each other in phase space. 
Here we show how this interference in phase space can be examined via
the factorisation of the analytic representation in the unit disk.
In particular we show how states whose analytic representation in the 
unit disk is a ``Blaschke factor'' can be produced as 
quantum superpositions.
 
We start by showing that a superposition of two inner states can be 
an outer state. It can be proved (see exercise 12 in chapter 5 of 
reference \cite{Hoff}) that if $Z(f;z)$ is an inner function then 
$Z(f;z) + 1$ is an outer function. A superposition of an inner state
$|f\rangle$ and the vacuum of the form $A(|f\rangle+|0\rangle)$ ($A$ 
is a normalisation constant) is represented by the analytic function 
$A[Z(f;z)+1]$ that is, by virtue of the above statement, an outer 
function (the multiplication by a constant leaves it outer).
Therefore, we can formulate the following theorem:
{\em the superposition of any inner state and the vacuum is an outer 
state}. For example, the superposition of the vacuum and any other
number state, $|n\rangle_{{\rm out}} = (|0\rangle + |n\rangle)/\sqrt{2}$, 
is an outer state because its analytic function $(1+z^{n})/\sqrt{2}$ is 
an outer function. All physical properties of such a state are determined 
by its phase distribution $(1+\cos n\theta)/2\pi$.

We also show how an inner state can be produced as a quantum 
superposition. We consider the superposition of the SU(1,1) coherent 
state $| z_0 \rangle$ (that is an outer state) and its first shifted 
state $| z_0 \rangle_{1} = \hat{E}_{+} | z_0 \rangle$ (these states will 
be studied in detail in section \ref{sec6.1}). This superposition is 
defined as
	\begin{equation}
| z_0 \rangle_{{\rm in}} = ( 1 - | z_0 |^2 )^{-1/2} ( | z_0 \rangle_{1}
- z_{0}^{\ast} | z_0 \rangle ) = (- z_{0}^{\ast}) |0\rangle +
( 1 - | z_0 |^2 ) \sum_{n=1}^{\infty} z_{0}^{n-1} |n\rangle . \label{zin}
	\end{equation}
The corresponding analytic function 
	\begin{equation}
Z^{({\rm in})}( z_0 ;z) = \frac{ z - z_0 }{ 1 - z_{0}^{\ast} z }
\label{fzin}
	\end{equation}
is a Blaschke factor \cite{Duren,Hoff} and we refer to the state 
(\ref{zin}) as a Blaschke state. It can be easily proved that the function
(\ref{fzin}) is an inner function. Therefore the Blaschke state (\ref{zin})
is an inner state, though it is composed of an outer state and a shifted 
outer state. In this case interference in phase space produces a state 
with a uniform phase distribution.

An interesting example is the superposition of the SU(1,1) coherent 
states $|z_0 \rangle$ and $|-z_0 \rangle$, defined as
	\begin{equation}
|z_0 ,\tau\rangle = {\cal N}^{-1/2} ( |z_0 \rangle + 
{\rm e}^{ {\rm i} \tau } |-z_0 \rangle )  \label{zsup}
	\end{equation}
where ${\cal N}$ is the normalisation factor,
	\begin{equation}
{\cal N} = 2 \left[ 1 + \frac{ 1 - | z_0 |^2 }{ 1 + | z_0 |^2 } 
\cos \tau \right] 
	\end{equation}
and $-\pi \leq \tau \leq \pi$.
The analytic function in the unit disk corresponding to the 
superposition state $|z_0 ,\tau\rangle$ is
	\begin{eqnarray}
& & Z(z_0 ,\tau;z) = {\cal A} \frac{ \cos(\tau/2) + {\rm i} \sin(\tau/2)
z_{0}^{\ast} z }{ 1 - z_{0}^{\ast 2} z^2 } \label{supfun} \\
& & {\cal A} \equiv 2 {\cal N}^{-1/2} (1-| z_0 |^2 )^{1/2} 
\exp(- {\rm i} \tau/2) .
	\end{eqnarray}
Important information about an analytic function can be obtained by
investigating its zeros. It is clear that any outer function has no 
zeros in the unit disk. An inner function can be written as a product
$S(z)B(z)$ where $S(z)$ is a function with no zeros (known as singular
function) and $B(z)$ is a function with zeros. The function $B(z)$ with 
zeros $\{\gamma_k \}$ can be written as a Blaschke product 
\cite{Duren,Hoff}:
	\begin{equation}
B(z) = \prod_k \left( \frac{ \gamma_{k}^{\ast} }{ |\gamma_k | }
\frac{ \gamma_k -z }{ 1- \gamma_{k}^{\ast} z } \right)^{ p_k } 
\label{Bla}
	\end{equation}
where $p_k$ are non-negative integers and $\gamma_k$ are distinct 
numbers in the unit disk. We see that the function $Z(z_0 ,\tau;z)$ of
equation (\ref{supfun}) has only one zero
	\begin{equation}
\gamma = \frac{ {\rm i} \cot(\tau/2) }{ z_{0}^{\ast} } \label{gamma}
	\end{equation}
that lies in the unit disk for $|z_0 | > |\cot(\tau/2)|$. We next
calculate the outer part of the function $Z(z_0 ,\tau;z)$ using
equation (\ref{3.3}). In the case $|z_0 | < |\cot(\tau/2)|$
(e.g. for $|\tau| \leq \pi/2$), we find $Z_{{\rm out}}(z_0 ,\tau;z) = 
Z(z_0 ,\tau;z)$; so the superposition state $|z_0 ,\tau\rangle$ with
$|z_0 | < |\cot(\tau/2)|$ is an outer state. This is consistent with
the fact that in this case the function $Z(z_0 ,\tau;z)$ has no zeros
in the unit disk. When $|z_0 | > |\cot(\tau/2)|$ (that is possible 
only for $|\tau| > \pi/2$), equation (\ref{3.3}) gives
	\begin{equation}
Z_{{\rm out}}(z_0 ,\tau;z) = {\cal A} \frac{ z_{0}^{\ast} }{ |z_0 | }
\frac{ z_0 \sin(\tau/2) + {\rm i} z \cos(\tau/2) }{ 
1 - z_{0}^{\ast 2} z^2 } .
	\end{equation}
Then the inner part is
	\begin{equation}
Z_{{\rm in}}(z_0 ,\tau;z) = \frac{ Z(z_0 ,\tau;z) }{ 
Z_{{\rm out}}(z_0 ,\tau;z) } = \frac{ \gamma^{\ast} }{ |\gamma| }
\frac{ \gamma -z }{ 1- \gamma^{\ast} z }  \label{blaf}
	\end{equation}
with $\gamma$ given by equation (\ref{gamma}). We see that in this case 
the inner part is exactly a Blaschke factor that has a zero in the unit
disk.

\section{Phase space for number and phase variables}
\label{sec6}

The purpose of this section is to demonstrate that the factorisation
into inner and outer parts is intimately connected to the phase-space 
formalism based on the number and phase variables. In section 
\ref{sec6.1} we study shifted states generated by the exponential phase
operator $\hat{E}_{+}$ and in section \ref{sec6.2} we consider the 
number-phase Wigner function.

\subsection{Shifted states}
\label{sec6.1}

We consider a state $|f\rangle$ with the analytic function
$Z(f;z)$ and define the state $|g\rangle$ as
        \begin{equation}
|g\rangle = \hat{W}(m,\beta,\gamma) |f\rangle  .   \label{3.27}
        \end{equation}  
It is easily seen that the analytic function $Z(g;z)$ of the state
$|g\rangle$ is given by
        \begin{eqnarray}
& & Z(g;z) = {\rm e}^{- {\rm i} \gamma} z^m 
Z(f; z {\rm e}^{- {\rm i} \beta})  \label{3.28} \\
& & Z_{{\rm in}}(g;z) = {\rm e}^{- {\rm i} \gamma} z^m
Z_{{\rm in}}(f; z {\rm e}^{- {\rm i} \beta})     \label{3.29} \\
& & \Phi(g;z) = \Phi(f; z {\rm e}^{- {\rm i} \beta}) .   \label{3.30}
        \end{eqnarray}
The corresponding boundary functions are related as
        \begin{eqnarray}
& & \Theta(g;\theta) = {\rm e}^{ {\rm i} m\theta - {\rm i} \gamma}
\Theta(f; \theta - \beta)   \label{3.31} \\
& & \Theta_{{\rm in}}(g;\theta) = {\rm e}^{ {\rm i} m\theta - 
{\rm i} \gamma} \Theta_{{\rm in}}(f; \theta - \beta)   \label{3.32} \\
& & \Phi(g;\theta) = \Phi(f; \theta - \beta) .   \label{3.33}
        \end{eqnarray}
When $\beta = 0$, the operator $\hat{W}$ changes only the inner part
of a state. Therefore, the phase distribution of a state is invariant
under the action of the operator $\hat{E}_{+}^{m}$. For an outer state
$|f\rangle$ with $Z(f;z) = Z_{{\rm out}}(f;z)$ we can define the
sequence $\{ |f\rangle_{m} \}$ of the shifted states
        \begin{equation}
|f\rangle_{m} = \hat{E}_{+}^{m} |f\rangle = \sum_{n=m}^{\infty}
f_{n-m} |n\rangle \;\;\;\;\;\;\;\;\;\; m=0,1,2,\ldots . \label{3.34}
        \end{equation}
All these states have identical outer functions $Z(f;z)$ and phase
distributions $P(f;\theta)$. The inner function of the shifted state
$|f\rangle_{m}$ is simply $z^m$. Therefore the number distribution is
shifted as
        \begin{equation}
P^{(m)}(f;n) = |\langle n|f \rangle_{m}|^{2} = \left\{
\begin{array}{l} P(f;n-m) \;\;\;\;\;\;\;\; n \geq m \\
0 \;\;\;\;\;\;\;\; n < m .   \end{array} \right. \label{3.35}
        \end{equation}
The Barut-Girardello representation of the shifted state 
$|f\rangle_{m}$ is given by
        \begin{equation}
U^{(m)}_{{\rm in}}(f;u) = \frac{u^m}{m!} \;\;\;\;\;\;\;\;\;\;
U^{(m)}(f;u) = \frac{1}{m!} \int_{0}^{u} {\rm d} x \, (u-x)^{m}
U_{{\rm out}}(f;x) .  \label{3.36}
        \end{equation}
Using equation (\ref{3.26}), we can express $U^{(m)}(f;u)$ with 
$m \geq 1$ as an integral over $U(f;u)$:
        \begin{equation}
U^{(m)}(f;u) = \frac{1}{(m-1)!} \int_{0}^{u} {\rm d} x \,
(u-x)^{m-1} U(f;x) .   \label{3.37}
        \end{equation}

For example, the SU(1,1) coherent state $|z\rangle$ generates the 
sequence $\{ |z\rangle_{m} \}$ of the shifted states
        \begin{equation}
|z\rangle_{m} = \hat{E}_{+}^{m} |z\rangle =
\sqrt{ 1 - | z |^2 } \sum_{n=m}^{\infty}
z^{n-m} |n\rangle \label{3.38}
        \end{equation}
which satisfy the eigenvalue equation
        \begin{equation}
\left( \hat{E}_{-} - |m-1 \rangle\langle m| \right) |z\rangle_{m} =
z |z\rangle_{m} . \label{3.39}
        \end{equation}
The Barut-Girardello state $|u\rangle$ generates the sequence
$\{ |u\rangle_{m} \}$ of the shifted states
        \begin{equation}
|u\rangle_{m} = \hat{E}_{+}^{m} |u\rangle = [I_{0}(2|u|)]^{-1/2}
\sum_{n=m}^{\infty} \frac{ u^{n-m} }{ (n-m)! }
|n\rangle \label{3.40}
        \end{equation}  
which satisfy the eigenvalue equation
        \begin{equation}
\hat{E}_{-} \left( \hat{N} - m \right) |u\rangle_{m} =
u |u\rangle_{m} . \label{3.41}
        \end{equation}
The states $|u\rangle_{m}$ have been recently considered as the
generalized philophase states \cite{Brif95}.

\subsection{The number-phase Wigner function}
\label{sec6.2}

For a quantum state $|f\rangle$ with the number representation $f_{n}$ 
and the phase representation $\Theta(f;\theta)$, the number-phase Wigner 
function is defined as \cite{Vacc}
	\begin{eqnarray}
S(f;n,\theta) & = & \frac{1}{2\pi} \left[ \sum_{p=-n}^{n} 
{\rm e}^{ {\rm i} 2p\theta} f_{n-p} f_{n+p}^{\ast} + \sum_{p=-n}^{n-1} 
{\rm e}^{ {\rm i} (2p+1)\theta} f_{n-p-1} f_{n+p}^{\ast} \right] 
\nonumber \\
& = & \frac{1}{(2\pi)^{2}} \int_{-\pi}^{\pi} {\rm d} \phi \,
(1+{\rm e}^{ {\rm i} \phi}) {\rm e}^{- {\rm i} 2n\phi} 
\Theta^{\ast}(f;\theta-\phi) \Theta(f;\theta+\phi) .    \label{3.42}
	\end{eqnarray}
The number and phase distributions are related to the number-phase 
Wigner function through the relations
	\begin{eqnarray}
& & P(f;n) = |\langle f|n \rangle|^{2} = 
\int_{-\pi}^{\pi} {\rm d} \theta \, S(f;n,\theta) \label{3.43} \\
& & P(f;\theta) = \frac{1}{2\pi} |\langle f|\theta \rangle|^{2}
= \sum_{n=0}^{\infty} S(f;n,\theta) . \label{3.44}
	\end{eqnarray}
The number-phase Wigner function of an outer state is uniquely
determined by its phase distribution $P(f;\theta)$. 
The number-phase Wigner function for the state $|g\rangle$ of 
equation (\ref{3.27}) is given by
	\begin{equation}
S(g;n,\theta) = \left\{ \begin{array}{l} S(f;n-m,\theta-\beta)  
\;\;\;\;\;\;\;\; n \geq m \\ 
0 \;\;\;\;\;\;\;\; n < m . \end{array} \right.  \label{3.45}
	\end{equation}

We calculate the number-phase Wigner function for some states 
discussed above. The number (Fock) state $|m\rangle$ (that is an 
inner state) is represented by the function
	\begin{equation}
S(m;n,\theta) = \frac{1}{2\pi} \delta_{n,m} .
	\end{equation}
For the superposition state 
$|m\rangle_{{\rm out}} = (|0\rangle +|m\rangle)/\sqrt{2}$
(that is an outer state), one obtains \cite{Vacc}
	\begin{equation}
S^{({\rm out})}(m;n,\theta) = \frac{1}{4\pi} [ \delta_{n,0} +
\delta_{n,m} + 2 \delta_{n,k} \cos(m\theta) ]
	\end{equation}
where $k=m/2$ if $m$ is even and $k=(m+1)/2$ if $m$ is odd.
The SU(1,1) coherent state $|z\rangle$ with 
$z = |z| {\rm e}^{ {\rm i} \phi }$ is represented by the function
	\begin{equation}
S(z;n,\theta) = \frac{ 1 - |z|^{2} }{ 2\pi } 
\left\{ |z|^{2n} U_{2n}[\cos(\theta-\phi)] + 
|z|^{2n-1}  U_{2n-1}[\cos(\theta-\phi)] \right\}   \label{3.46}
	\end{equation}
where $U_{n}(x)$ is the $n$-order Chebyshev polynomial of the second
kind:
	\begin{equation}
U_{n}(\cos \theta) = \frac{ \sin (n+1)\theta }{ \sin \theta }
	\end{equation}
and $U_{n}(x) \equiv 0$ for $n < 0$.
For the Barut-Girardello state $|u\rangle$ 
with $u = |u| {\rm e}^{ {\rm i} \varphi }$, we find
	\begin{equation}
S(u;n,\theta) = \frac{1}{2\pi I_{0}(2|u|)} \left\{  
\frac{ [2|u|\cos(\theta-\varphi)]^{2n} }{ (2n)! } + 
\frac{ [2|u|\cos(\theta-\varphi)]^{2n-1} }{ (2n-1)! }
\right\} .  \label{3.47}
	\end{equation}
The second term in equations (\ref{3.46}), (\ref{3.47}) should be 
omitted for $n=0$. The number-phase Wigner functions for the shifted 
states $|z\rangle_{m}$ and $|u\rangle_{m}$ are obtained from equations 
(\ref{3.46}) and (\ref{3.47}), respectively, according to the
relation (\ref{3.45}).
For the superposition state $|z\rangle_{{\rm in}}$ of equation 
(\ref{zin}) (that is an inner state) the number-phase Wigner function 
can be written in terms of the Chebyshev polynomials: 
	\begin{eqnarray}
\hspace{-0.8cm}
S^{({\rm in})}(z;n,\theta) & = & \frac{1}{2\pi} \left\{ 
r^{2n-3} U_{2n-3}[\cos(\theta-\phi)] + [1-2r\cos(\theta-\phi)]
r^{2n-2} U_{2n-2}[\cos(\theta-\phi)] \right. \nonumber \\
& & \left. + [r^2 -2r\cos(\theta-\phi)]
r^{2n-1} U_{2n-1}[\cos(\theta-\phi)] + 
r^{2n+2} U_{2n}[\cos(\theta-\phi)] \right\} 
	\end{eqnarray}
where $z=r {\rm e}^{ {\rm i} \phi}$. As we have shown, the 
superposition state $|z,\tau\rangle$ of equation (\ref{zsup}) is an
outer state for $|z|<|\cot(\tau/2)|$, while for $|z|>|\cot(\tau/2)|$ 
it has an inner part that is a Blaschke factor (see equation 
(\ref{blaf})). The state $|z,\tau\rangle$ with 
$z=r {\rm e}^{ {\rm i} \phi}$ is represented by the number-phase 
Wigner function
	\begin{eqnarray}
S(z,\tau;n,\theta) & = & \frac{ |{\cal A}|^{2} }{ 2\pi } \left\{ 
[\cos^{2}(\tau/2)] r^{2n} U_{n}[\cos 2(\theta-\phi)] \right. \nonumber \\
& & \left. + [r^{2} \sin^{2}(\tau/2) - r\sin(\tau) \sin(\theta-\phi)]
r^{2(n-1)} U_{n-1}[\cos 2(\theta-\phi)] \right\} .
	\end{eqnarray}
For $r<|\cot(\tau/2)|$ this function is uniquely
determined by its marginal distribution $P(z,\tau;\theta)$; while for
$r>|\cot(\tau/2)|$ there exists the inner part that brings new features
which cannot be described by the phase distribution alone.

\section{Conclusions}
\label{sec:concl}

We have shown that the factorisation of analytic representations 
in the unit disk into inner and outer parts has important physical
significance. It is intimately related to the number-phase Weyl 
semigroup and the number-phase quantum statistical properties.
This factorisation can be implemented not only in analytic 
representations in the unit disk, but also in other representations.
 
We have given several examples of inner and outer states and we have
constructed explicitly Blaschke states. Inner states have a uniform 
phase distribution while outer states are uniquely determined by their 
phase properties. This work has demonstrated that
pure mathematical concepts in the theory of Hardy spaces have an
interesting physical interpretation.

\section*{Acknowledgements}

AV gratefully acknowledges support from the British council in 
the form of a travel grant.
CB gratefully acknowledges the financial help from the Technion.
AM was supported by the Fund for Promotion of Research
at the Technion, by the Technion -- VPR Fund, and by GIF --- 
German-Israeli Foundation for Research and Development.

\end{document}